\documentclass[12pt]{article}
\usepackage{graphicx}
\usepackage{graphicx}
\newlength{\vshift}
\input epsf.tex
\newlength{\hshift}

\usepackage{amssymb}
\usepackage{amsmath}
\usepackage{amsthm}
\usepackage{amsopn}
\def\uno{\mbox{1 \kern-.59em {\rm l}}}

\def\beq{\begin{equation}}
\def\eeq{\end{equation}}
\def\bea{\begin{eqnarray}}
\def\eea{\end{eqnarray}}

\begin{document}

 \vspace*{3cm}

\begin{center}

{\bf{\Large Thermodynamic Geometry of Fractional Statistics}}

\vskip 4em

{ {\bf Behrouz ~Mirza} \footnote{e-mail: b.mirza@cc.iut.ac.ir}\:
and \: {\bf Hosein ~Mohammadzadeh}\footnote{e-mail:
h.mohammadzadeh@ph.iut.ac.ir}}

\vskip 1em

Department of Physics, Isfahan University of Technology, Isfahan,
84156-83111, Iran

 \end{center}

 \vspace*{1.9cm}

\begin{abstract}
We extend our earlier study about the fractional exclusion statistics to
higher dimensions in full physical range and in the non-relativistic and ultra-relativistic limits.
Also, two other fractional statistics, namely Gentile and Polychronakos fractional statistics, will
be considered and similarities and differences between these statistics will be
explored. Thermodynamic geometry suggests that a  two dimensional  Haldane fractional exclusion gas is more stable than
higher dimensional gases. Also, a complete picture of attractive and repulsive statistical interaction of
fractional statistics is given. For a special kind of fractional
statistics, by considering the singular points of thermodynamic
curvature, we find a condensation for a non-pure bosonic system
which is similar to the Bose-Einstein condensation and the phase transition temperature will be worked out.
\end{abstract}

PACS number(s): 05.20.-y, 67.10.Fj
\newpage
\section{Introduction}
As a generalization of Bose-Einstein and Fermi-Dirac statistics,
 fractional statistics has been discussed for many years \cite{Leinas, Goldin,wilczek,medvedov}.
The idea of generalized Pauli principle or fractional exclusion
statistics was introduced by Haldane \cite{Haldane}. His
definition of particles obeying the generalized exclusion
statistics with finite dimensional Hilbert space is motivated by
such physical examples, as quasi particle in the fractional
quantum Hall system and spinons in anti-ferromagnetic spin chains.
The parameter governing fractional exclusion statistics in this
case is defined by $g=-\frac{\Delta d}{\Delta N}$, where $\Delta
d$ is the change in the dimension of the single particle Hilbert
space and $\Delta N$ is the change in the number of particles
when the size of the system and the boundary conditions are
unchanged. By definition, therefore, $g=0$ corresponds to bosons
and $g=1$ to fermions. A different concept of fractional
statistics, namely, fractional exchange statistics arises when the
many body wave function of a system of indistinguishable
particles is allowed to acquire an arbitrary phase
$e^{i\pi\alpha}$ upon an adiabatic exchange process of two
particles. Here, $\alpha$ is the so-called exchange statistical
parameter, interpolating between $\alpha=0$ (bosons) and
$\alpha=1$ (fermions). Such an exchange produces a nontrivial
phase only if the configuration space of the collection of
particles under study possesses a multiply connected topological
structure. Therefore, fractional exchange statistics is usually
restricted to two spatial dimensions and can also be formalized,
to some extent, in one spatial dimension. However, fractional
exclusion statistics, is based on the structure of the Hilbert
space, rather than on the configuration space, of the particle
assembly and is, thus, not restricted to $d\leq 2$. In many
systems, it is also possible to relate the exclusion statistics
parameter, $g$, to the exchange statistics parameter, $\alpha$.
The thermodynamic properties of two sorts of fractional statistics
have been considered by several authors
\cite{nayak,murthy1,murthy2,wilczek1}. Specially, the
thermodynamics of the anyon gas---a two dimensional system with
fractional statistics---has been the subject of much research
\cite{Wu,wung1,wung2,wung3,pellegrino,batchelor}. More recently
the thermodynamic geometry of the anyon gas has been considered
\cite{Mirza1,Mirza2}. There is also another sort of fractional
exclusion statistics, introduced by Polychronakos\cite{poly}. It
is mentionable that the same distribution function was was
derived from collision theory by using the detailed balance
hypothesis \cite{March1,March2}. Furthermore, Gentile statistics
is a generalization of Fermi-Dirac and Bose-Einstein statistics
which is based on allowing  $p$ ($p\in[1,\infty[$) particles to
occupy the same quantum states \cite{Gentile1,Gentile2}.

The geometrical structure of the phase space of statistical
thermodynamics has been studied and the thermodynamic curvature
has already been calculated for some models whose thermodynamics
is exactly known, where reviews for these models can be found in
\cite{Ruppeiner01} and \cite{brody2}. This approach has been
utilized to study the thermodynamics of black holes
\cite{aman,mirza,Ruppeiner,alvarez,sahay,banerjee}. The
thermodynamic curvature of the ideal classical gas is zero and it
could be a criterion for statistical interaction of the system
\cite{Ruppeiner1,Nulton}. Janyszek and Mruga{\l}a worked out the
thermodynamic curvature for ideal Fermi and Bose gases and
reported that the sign of the thermodynamic curvature is always
different for ideal Fermi and Bose gases. It may be shown that the
sign of thermodynamic curvature specifies the attractive or
repulsive statistical interaction of systems.  It has been argued
that the scalar curvature could be used to show that fermion
gases are  more stable than boson gases \cite{Mrugala2}. Also,
phase transition properties of van der Waals gas and some other
thermodynamic models have been considered and it has been shown
that the singular point of the thermodynamic curvature coincides
with the critical point of the system \cite{brody,Janke}.
Recently the thermodynamic curvature of the classical limit of the
anyon gas has been worked out \cite{Mirza1}. Also, the full
physical range thermodynamic curvature of the  anyon gas has been
obtained by using its factorizable property \cite{Mirza2}.
However, these  properties do not hold in arbitrary dimensions
\cite{wung1,wung2,wung3}.

The outline of this paper is as follows. In Sec. 2, the
thermodynamic properties of particles obeying Haldane fractional
exclusion (HFE) statistics are summarized and the internal energy
and the particle number in an arbitrary  dimension $D$ is derived.
Also, the same thermodynamic quantities are obtained for
Polychronakos fractional exclusion (PFE) statistics and Gentile
statistics. In Sec. 3, the thermodynamic geometry of three kinds
of fractional statistics is investigated and some interesting
properties of these systems are explored. Finally, in Sec. 4,
condensation of particles with fractional statistics is explored.

\section{The ideal gas of fractional statistical particles }
\subsection{Haldane fractional exclusion statistics}
Fractional exclusion statistics was introduced as a generalization
of Pauli exclusion for fermions. Haldane defined the exclusion
statistics parameter, $g$, of a particle by
 \bea
 g=-\frac{d_{N+\Delta N}-d_{N}}{\Delta N},
 \eea
where $N$ is the number of particles and $d_N$ is the dimension of
one particle Hilbert space obtained by holding the coordinate of
$N-1$ particles fixed \cite{Haldane}. The statistical distribution  function, $n(\zeta)$,  of an ideal gas of fractional statistical particles with the chemical potential $\mu$ and temperature $T$ has been obtained by Wu using the Haldane's fractional statistics \cite{Wu},
    \bea
    \label{n}
    n(\zeta)=\frac{1}{\textit{w}(e^{(\epsilon-\mu)/kT})+g},
    \eea
where, the function $\textit{w}(\zeta)$ satisfies the functional
equation
    \bea\label{w}
    \textit{w}(\zeta)^{g}[1+\textit{w}(\zeta)]^{1-g}=\zeta\equiv
    e^{(\epsilon-\mu)/kT}.
    \eea
The functional equation for $\textit{w}(\zeta)$ can be solved
analytically only in a few special cases. Equation (\ref{w})
yields the correct solutions for two familiar cases: bosons
$(g=0)$, $\textit{w}(\zeta)=\zeta-1$ and fermions $(g=1)$,
$\textit{w}(\zeta)=\zeta$. In the thermodynamic limit, the
internal energy and particle number of an exclusion gas in a $D$
dimensional box of volume $L^D$ with the following dispersion
relation
 \bea
 \epsilon=ap^\sigma
 \eea
can be written as
 \bea
 U&=&\int_{0}^{\infty}\epsilon \
 n(\zeta)\Omega(\epsilon)d\epsilon\nonumber,\\
 N&=&\int_{0}^{\infty}n(\zeta)\Omega(\epsilon)d\epsilon\label{pn},
 \eea
and $\Omega(\epsilon)$ is the density of the single particle state
for the system. Neglecting the spin of particle, the standard form
of density of states will be
 \bea
 \Omega(\epsilon)=\frac{A^D}{\Gamma(\frac{D}{2})}\epsilon^{D/\sigma
 -1},
 \eea
where, $\sigma$ takes some values for example $(\sigma=2)$ for
non relativistic and $(\sigma=1)$ for ultra relativistic
particles and $A=\frac{L\sqrt{\pi}}{a^{1/\sigma} h}$ is a
constant and for simplicity we will set it equal to one $(A=1)$.
Obtaining  the internal energy and particle number of the ideal
exclusion gas for special cases where  Eq. (\ref{w}) could be
solved will be straightforward. It is possible to solve this
equation using elementary methods for special cases
$g=\frac{1}{4},\frac{1}{3},\frac{1}{2},\frac{2}{3},\frac{3}{4},$
\cite{Aoyama}. Also, an analytical explicit series solution for
the general values of $g$ has been presented by using Lagrange
inversion theorem \cite{Geoffrey}. It was shown that there is a
factorizable property for thermodynamic quantities of a two
dimensional HFE gas that indicates we can evaluate the internal
energy and particle number of the anyon gas as a composition of
the internal energy and particle number of fermion and boson
gases, while the particle number of anyon, fermion, and boson
gases are the same. Moreover, it has been argued that the
factorizable property holds only for two spatial dimensions
\cite{wung3}. Therefore, the internal energy and particle number
cannot be evaluated analytically   for the general value of
fractional exclusion parameter in an arbitrary dimension.
However, another method is available that enables us to explore
the thermodynamic geometry of an ideal gas with fractional
statistics in arbitrary $D$ dimensions \cite{potter}. First, we
change the integrating variable $\epsilon$ to $\textit{w}$. Using
Eq. (\ref{w}), we will have
 \bea
 \beta(\epsilon-\mu)&=&g\ln(\textit{w}) + (1- g)\ln(\textit{w} + 1),\nonumber\\
 d\epsilon&=&\frac{1}{\beta}\frac{\textit{w}+g}{\textit{w}(1+\textit{w})}d\textit{w}.
 \eea
We then have
 \bea
 U=\frac{A^D}{\Gamma(\frac{D}{2})}\beta^{-\frac{D}{\sigma}-1}\int_{\textit{w}_{0}}^{\infty}
 \frac{\left\{\ln(z\textit{w}^{\  g}\ (1+\textit{w})^{1-g})\right\}^{\frac{D}{\sigma}}}{\textit{w}(1+\textit{w})}d\textit{w},\nonumber\\
 N=\frac{A^D}{\Gamma(\frac{D}{2})}\beta^{-\frac{D}{\sigma}}\int_{\textit{w}_{0}}^{\infty}
 \frac{\left\{\ln(z\textit{w}^{\ g}\ (1+\textit{w})^{1-g})\right\}^{\frac{D}{\sigma}-1}}{\textit{w}(1+\textit{w})}d\textit{w},\label{UNH}
 \eea
where, $\textit{w}_{0}$ corresponds to the value of $\textit{w}$
for $\epsilon=0$, which satisfies relation (\ref{bm})
 \bea
 -\beta\mu=g\ln(\textit{w}_{0}) + (1- g)\ln(\textit{w}_{0}+ 1)\label{bm}
 \eea
and $z=e^{\mu/kT}$ is the fugacity of gas. Therefore, for an
arbitrary value of fractional exclusion parameter, $g$,
temperature $T=1/\beta k_{_B}$, and fugacity $z$ of the system,
the proper value of $\textit{w}_{0}$ could be evaluated and the
value of internal energy and particle number will be determined.
\subsection{Gentile statistics}
Gentile generalized the Fermi-Dirac and Bose-Einstein statistics
by allowing for the possibility of (no more than) $p$ particles to
occupy the same quantum state and then derived the following
equation for the distribution function
\cite{Gentile1,Gentile2,khare}:
 \bea
  n=\frac{1}{e^{(\epsilon-\mu)/kT}-1}-\frac{p+1}{e^{p+1(\epsilon-\mu)/kT}-1},
 \eea
where, for $p=1$, one obtains the Fermi-Dirac distribution
function and when $p$ goes to infinity, the Bose-Einstein
distribution function will be acquired. We will set
$q=\frac{1}{p}$, then $q=0$ will correspond to the boson and $q=1$
to the fermion case. Then, for $0<q<1$, there is a new kind of
fractional statistics, namely Gentile statistics. It is
straightforward to obtain the internal energy and particle number
of an ideal gas with particles obeying Gentile statistics,
 \bea
 U&=&A^{D}\frac{\Gamma(\frac{D}{\sigma}+1)}{\Gamma(\frac{D}{2})}\beta^{-(\frac{D}{\sigma}+1)}[Li_{\frac{D}{\sigma}+1}(z)\nonumber\\
 &-&(\frac{q+1}{q})^{-\frac{D}{\sigma}}
 Li_{\frac{D}{\sigma}+1}(z^{\frac{q+1}{q}})],\nonumber\\
 N&=&A^{D}\frac{\Gamma(\frac{D}{\sigma})}{\Gamma(\frac{D}{2})}\beta^{-(\frac{D}{\sigma})}[Li_{\frac{D}{\sigma}}(z)\nonumber\\
 &-&(\frac{q+1}{q})^{-(\frac{D}{\sigma}-1)}
 Li_{\frac{D}{\sigma}}(z^{\frac{q+1}{q}})],\label{UNG}
 \eea
where, $Li_{n}(x)$ denotes the polylogarithm function.

\subsection{Polychronakos fractional statistics}
Polychronakos  suggested another kind of fractional exclusion
 statistics \cite{poly}. It can be thought of as a different realization
 of the exclusion statistics idea. Unlike the Haldane fractional exclusion
 statistics, the distribution function has a simple form  given by
 \bea
 n=\frac{1}{e^{(\epsilon-\mu)/kT}+\alpha}.
 \eea
Fermions and bosons correspond to $\alpha=1$ and $\alpha=-1$,
respectively, while $\alpha=0$ corresponds to Boltzmann
statistics. By substituting $\alpha=2k-1$, fermions and bosons
will correspond
 to $k=0$ and $k=1$ and Polychronakos fractional statistics corresponds to $0<k<1$. The internal energy and particle
 number are given by:
 \bea
 U&=&A^{D}\ \frac{\Gamma(\frac{D}{\sigma}+1)}{\Gamma(\frac{D}{2})}\beta^{-(\frac{D}{\sigma}+1)}\ \frac{Li_{\frac{D}{\sigma}+1}(z-2kz)}{1-2k},\nonumber\\
  N&=&A^{D}\ \frac{\Gamma(\frac{D}{\sigma})}{\Gamma(\frac{D}{2})}\beta^{-(\frac{D}{\sigma})}\ \frac{Li_{\frac{D}{\sigma}}(z-2kz)}{1-2k}.\label{UNP}
 \eea

\section{Thermodynamic curvature of fractional statistic gas}
The geometrical structure of the phase space of statistical
thermodynamics was explicitly studied by Gibbs. The geometrical
thermodynamics was developed by Ruppeiner and Wienhold
\cite{Ruppeiner1,weinhold}. Ruppeiner geometry is based on the
entropy representation, where we denote the extended set of $n+1$
extensive variables of the system by
$X=(U,N^{1},...,V,...,N^{r})$, while Weinhold worked in the
energy representation in which the extended set of $n+1$
extensive variables of the system are denoted by
$Y=(S,N^{1},...,V,...,N^{r})$ \cite{Ruppeiner01}. It should be
noted that we can work in any thermodynamic potential
representation that is the Legendre transform of the entropy or
the internal energy. The metric of this representation may be the
second derivative of the thermodynamic potential with respect
 to intensive variables; for example, the thermodynamic potential
$\Phi$ which is defined as,
    \bea
    \Phi=\Phi(\{F^{i}\})=\Phi(1/T,-\mu^{1}/T,\ldots,P/T,\ldots,-\mu^{r}/T),
    \eea
$\Phi$ is the
Legendre transform of entropy with respect to the extensive
parameter, $X^{i}$,
    \bea
    F^{i}=\frac{\partial S}{\partial X^{i}},
    \eea
and $r$ identify the number of various kind of particles.
The metric in this representation is given by
    \bea
    g_{ij}=\frac{\partial^{2}\Phi}{\partial F^{i}\partial F^{j}}.
    \eea
Janyszek and Mruga{\l}a used the partition function to introduce
the metric geometry of the parameter space \cite{Mrugala2},
    \bea\label{M1} g_{ij}=\frac{\partial^{2}\ln
    Z}{\partial \beta^{i}\partial\beta^{j}},
    \eea
where, $\beta^{i}=F^{i}/k$ and $Z$ is the partition function.

\subsection{Haldane fractional exclusion statistcs}

According to Eq. (\ref{UNH}), the parameter space of an ideal
fractional exclusion gas is  $(\beta,\gamma)$, where $ {\beta} =
{1 / kT}$ and $\gamma = -\mu/ kT $ . For computing the
thermodynamic metric, we select one of the extended variables as
the constant system scale. We will implicitly pick Volume in
working with the grand canonical ensemble \cite{Mrugala2}. We can
evaluate the metric elements by the definition of metric in Eq.
({\ref{M1}). The metric elements of the thermodynamic space of an
ideal HFE gas are given by
    \bea
     \label{vv} &&G_{\beta\beta}=\frac{\partial^{2}\ln Z}{\partial
    \beta^{2}}=-(\frac{\partial U}{\partial\beta})_{\gamma}\nonumber\\
   &&=\frac{-(\frac{D}{\sigma} +1)\beta^{-(\frac{D}{\sigma}+2)}}{\Gamma(\frac{D}{2})}\int_{\textit{w}_{0}}^{\infty}
 \frac{\left\{\ln(z\textit{w}^{\ g}\ (1+\textit{w})^{1-g})\right\}^{\frac{D}{\sigma}}}{\textit{w}(1+\textit{w})}d\textit{w},\nonumber\\
    &&G_{\beta\gamma}=\frac{\partial^{2}\ln Z}{\partial\gamma\partial\beta}=-(\frac{\partial N}{\partial \beta})_{\gamma}\nonumber\\
    &&=\frac{-\frac{D}{\sigma}\beta^{-(\frac{D}{\sigma}+1)} }{\Gamma(\frac{D}{2})}\int_{\textit{w}_{0}}^{\infty}
 \frac{\left\{\ln(z\textit{w}^{\ g}\ (1+\textit{w})^{1-g})\right\}^{\frac{D}{\sigma}-1}}{\textit{w}(1+\textit{w})}d\textit{w},\nonumber\\
     &&G_{\gamma\gamma}=\frac{\partial^{2}\ln Z}{\partial\gamma^{2}}=-(\frac{\partial N}{\partial\gamma})_{\beta}\nonumber\\
     &&=\frac{-\beta^{-\frac{D}{\sigma}}}{\Gamma(\frac{D}{2})}\int_{\textit{w}_{0}}^{\infty}
 \frac{\left\{\ln(z\textit{w}^{\ g}\ (1+\textit{w})^{1-g})\right\}^{\frac{D}{\sigma}-1}}{(\textit{w}+g)^{2}}d\textit{w}.
    \eea
We note that the value of $\textit{w}_{0}$, according to Eq.
(\ref{bm}), is independent of $\beta$ and depend only on
$\gamma$. Therefore, in differentiating with respect to $\beta$,
$\textit{w}_{0}$ is a constant. To obtain the last equation, Eq.
(\ref{pn}) is used and differentiated with respect to
$\textit{w}$. $\frac{\partial\textit{w}}{\partial\gamma}$ is
evaluated  from Eq. (\ref{bm}). \bea
&&(\frac{\partial N}{\partial\gamma})_{\beta}=\frac{\partial N}{\partial\textit{w}}\frac{\partial\textit{w}}{\partial\gamma}\nonumber\\
&&=\int_{0}^{\infty}\frac{-\Omega(\epsilon)}{(\textit{w}+g)^2}\frac{e^{(\epsilon-\mu)/kT}\textit{w}^{\
g}\ (1+\textit{w})^{1-g}}{\textit{w}+g}d\epsilon\label{nk}. \eea
By changing the integrating variable to $\textit{w}$, the equation
will be obtained.
We consider a system with two thermodynamic degrees of freedom
and,
 therefore, the dimension of the thermodynamic surface or parameter
space is equal to two. Thus, the scalar curvature is
given by
    \bea
    R=\frac{2}{\det g} R_{1212.}
    \eea
 Janyszek and Mruga{\l}a \cite{Mrugala3} demonstrated
 that if the metric elements  are written purely as
the second derivatives of a certain thermodynamic potential, the
thermodynamic curvature may then be written in terms of the second
and the third derivatives. The sign convention for $R$ is
arbitrary, so $R$ may be either negative or positive for any
case. Our selected sign convention is the same as that of Janyszek
and Mruga{\l}a, but different from \cite{Ruppeiner01}. In two
dimensional thermodynamic spaces, the Ricci scalar is defined by,

 \bea
R=\frac{2\left|
         \begin{array}{ccc}
           g_{\beta\beta} & g_{\gamma\gamma} & g_{\beta\gamma} \\
           g_{\beta\beta,\beta} & g_{\gamma\gamma,\beta} & g_{\beta\gamma,\beta} \\
           g_{\beta\beta,\gamma} & g_{\gamma\gamma,\gamma} & g_{\beta\gamma,\gamma} \\
         \end{array}
       \right|}{{\left|
                  \begin{array}{cc}
                    g_{\beta\beta} & g_{\beta\gamma} \\
                    g_{\beta\gamma} & g_{\gamma\gamma} \\
                  \end{array}
                \right|}^{2}
       }.
    \eea
Using the following equations for an ideal fractional exclusion gas
    \bea
    G_{\beta\beta,\beta}&=&\frac{(\frac{D}{\sigma} +1)(\frac{D}{\sigma} +2)}{\Gamma(\frac{D}{2})}\beta^{-(\frac{D}{\sigma}+3)}\nonumber\\
    &\times&\int_{\textit{w}_{0}}^{\infty}
 \frac{\left\{\ln(z\textit{w}^{g}(1+\textit{w})^{1-g})\right\}^{\frac{D}{\sigma}}}{\textit{w}(1+\textit{w})}d\textit{w},\nonumber\\
    G_{\beta\beta,\gamma}&=&G_{\beta\gamma,\beta}=\frac{(\frac{D}{\sigma})(\frac{D}{\sigma}+1) }{\Gamma(\frac{D}{2})}\beta^{-(\frac{D}{\sigma}+2)}\nonumber\\
    &\times&\int_{\textit{w}_{0}}^{\infty}
 \frac{\left\{\ln(z\textit{w}^{g}(1+\textit{w})^{1-g})\right\}^{\frac{D}{\sigma}-1}}{\textit{w}(1+\textit{w})}d\textit{w},\nonumber\\
   G_{\gamma\gamma,\beta}&=&G_{\beta\gamma,\gamma}=\frac{\frac{D}{\sigma} }{\Gamma(\frac{D}{2})}\beta^{-(\frac{D}{\sigma}+1)}\nonumber\\
   &\times&\int_{\textit{w}_{0}}^{\infty}
 \frac{\left\{\ln(z\textit{w}^{g}(1+\textit{w})^{1-g})\right\}^{\frac{D}{\sigma}-1}}{(\textit{w}+g)^{2}}d\textit{w},\\
    G_{\gamma\gamma,\gamma}&=&\frac{1}{\Gamma(\frac{D}{2})}\beta^{-\frac{D}{\sigma}}\int_{\textit{w}_{0}}^{\infty}
 \{\textit{w}(\textit{w}+2)-g(1+2\textit{w})\}\nonumber\\
 &\times&\frac{\left\{\ln(z\textit{w}^{g}(1+\textit{w})^{1-g})\right\}^{\frac{D}{\sigma}-1}}
 {(\textit{w}+g)^{4}}d\textit{w}.\nonumber
    \eea

The last equation is obtained by applying the chain rule from Eq.
(\ref{nk}) and differentiating with respect to $\gamma$. Now,
everything required for evaluating the thermodynamic curvature is
available. We will focus on two non-relativistic and
ultra-relativistic limits.

\subsubsection{Non-Relativistic Limit ($\sigma=2$)}
Now, we can work out the thermodynamic curvature for an ideal HFE
gas in an arbitrary dimension. Specially, we are interested in
investigating the behavior of the thermodynamic curvature in three
spatial dimensions. The thermodynamic geometry of fractional
exclusion gas in two dimensions has completely been considered
based on the factorizable property. The thermodynamic curvature
for isothermal processes and some fixed
 fugacity values as a function of fractional parameter in the non relativistic
 limit ($\sigma=2$) are depicted in Figs. \ref{figure1} and \ref{figure2}.
   \begin{figure}[t]
    \center
    \includegraphics[width=1.1\columnwidth]{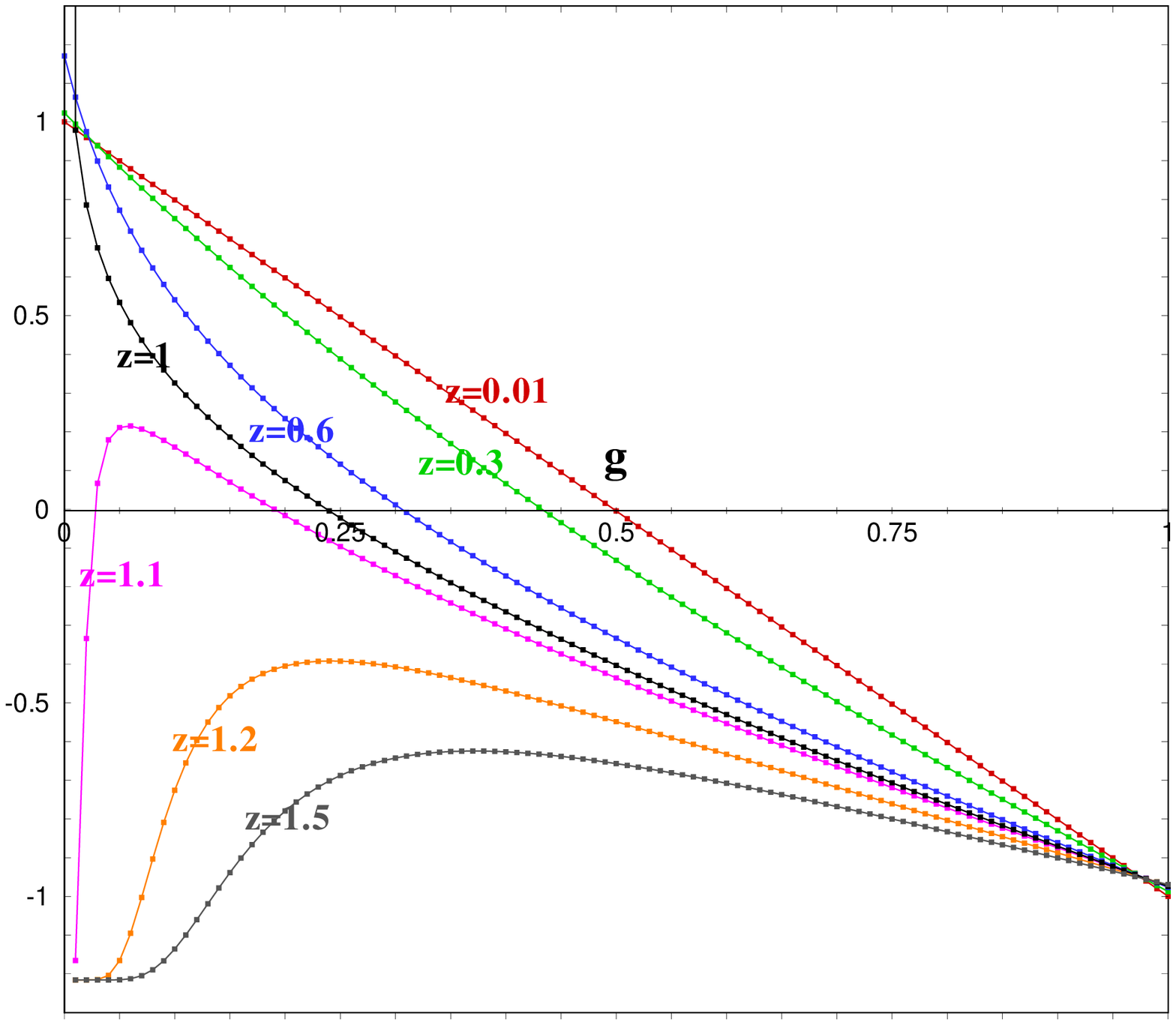}\\
    \caption{(Color online) \ The thermodynamic curvature of a 2D  HFE gas (non-relativistic) with
    respect to the fractional parameter, g, for an isotherm ($\beta=1$). The assumed values of anyon fugacity
      are  $ z=0.01$  [red (upper) line],  0.3  [green line],
     0.6 (blue line), 1 [black (middle) line], 1.1 (purple line), 1.2 (orange line),
      1.5 [light gray (lower) line].}\label{figure1}
   \end{figure}

   \begin{figure}[t]
    \center\includegraphics[width=1.1\columnwidth]{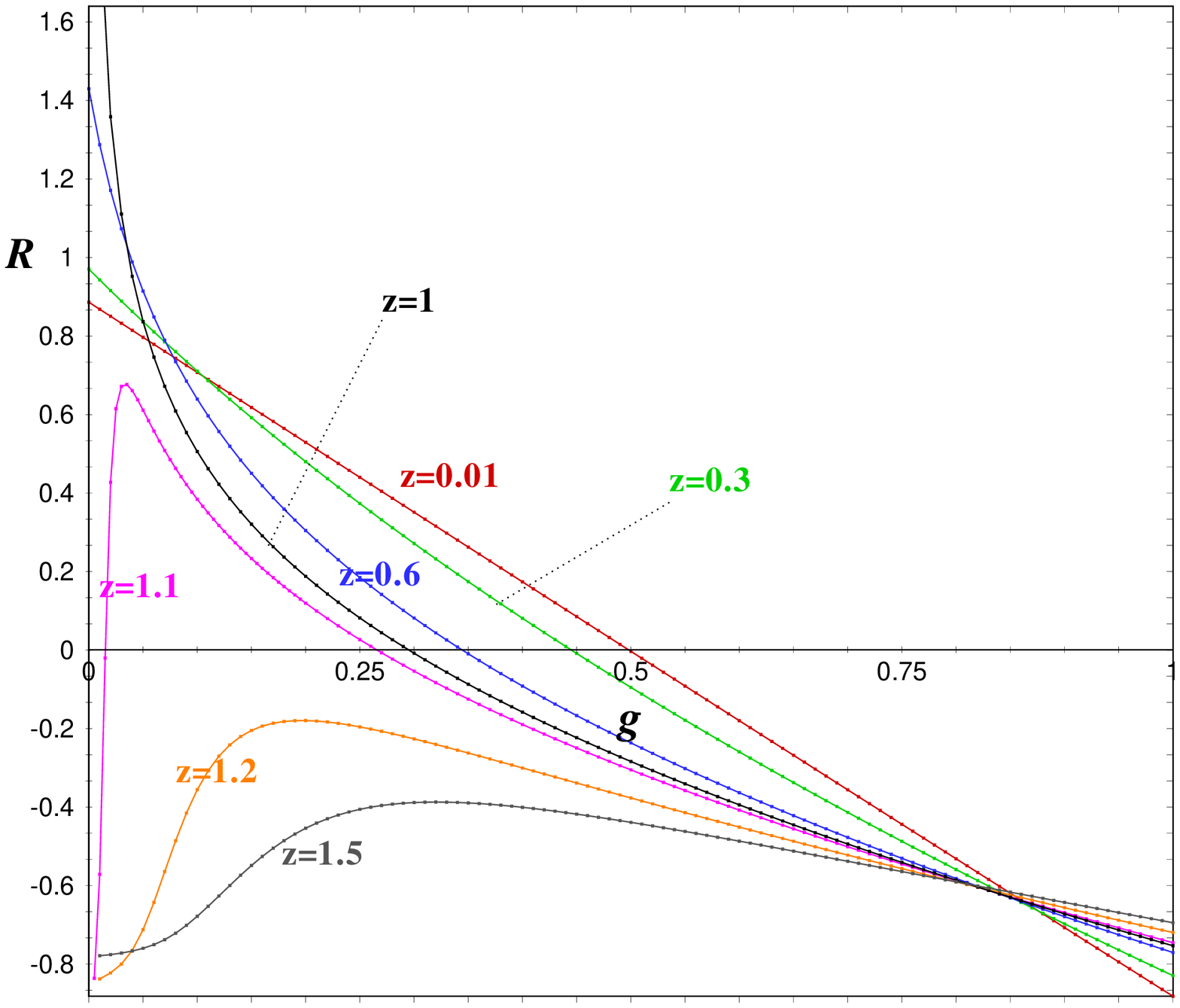}\\
    \caption{(Color online)  The thermodynamic curvature of a 3D  HFE gas  with respect to the fractional
     parameter for an isotherm ($\beta=1$). The assumed values of fugacity
     are  $ z=0.01$ [red (upper) line], 0.3  [green line],
     0.6 (blue line), 1 [black (middle) line], 1.1 (purple line), 1.2 (orange line),
      1.5 [light gray (lower) line].}\label{figure2}
   \end{figure}

\noindent Obviously, the thermodynamic curvature for two
dimensional gases based on this method coincides with the results
of our previous paper \cite{Mirza2}. Also, the result for a three
dimensional fractional exclusion gas is similar to those of the
two dimensional case. The result is expandable to an arbitrary
dimension. In the non-relativistic limit, it seems that in all
dimensions and in the high temperature (classical or Boltzmann)
limit, the sign of the thermodynamic curvature changes at the same
value of fractional parameter $g=0.5$. Therefor, in the high
temperature limit and in all dimensions, the fractional exclusion
gas is divided in to two cases. For $g<0.5$, the thermodynamic
curvature is positive or "Bose-like" and for $g>0.5$, it is
negative or "Fermi-like". For $g=0.5$, the ideal HFE gas behaves
such as an ideal classical gas. Also, one can find an interesting
behavior for $z>1$. The thermodynamic curvature has a maximum
point. According to the stability interpretation of the value of
the  thermodynamic curvature  these maximum points should be
related to a less stable state of the system. It has been shown
\cite{Mrugala2,janyszek} that we may consider the thermodynamic
curvature as a measure of the stability of the system: The bigger
the value of R, the less stable is the system. This
interpretation of stability measures the looseness of the system
to fluctuations and does not refer to the fact that the metric is
definitely positive. It is also interesting to note that we  can
find two different values of $g$ with the same value for the
thermodynamic curvature. For some values of $z$, we obtain two
values for $g$ with zero curvature, which indicates a duality
relation between these points. The thermodynamic curvature of a
three dimensional fractional exclusion
 gas is depicted in Fig. \ref{figure3} for several values of fractional parameter as a function of fugacity.
\begin{figure}[t]
    \center\includegraphics[width=1.1\columnwidth]{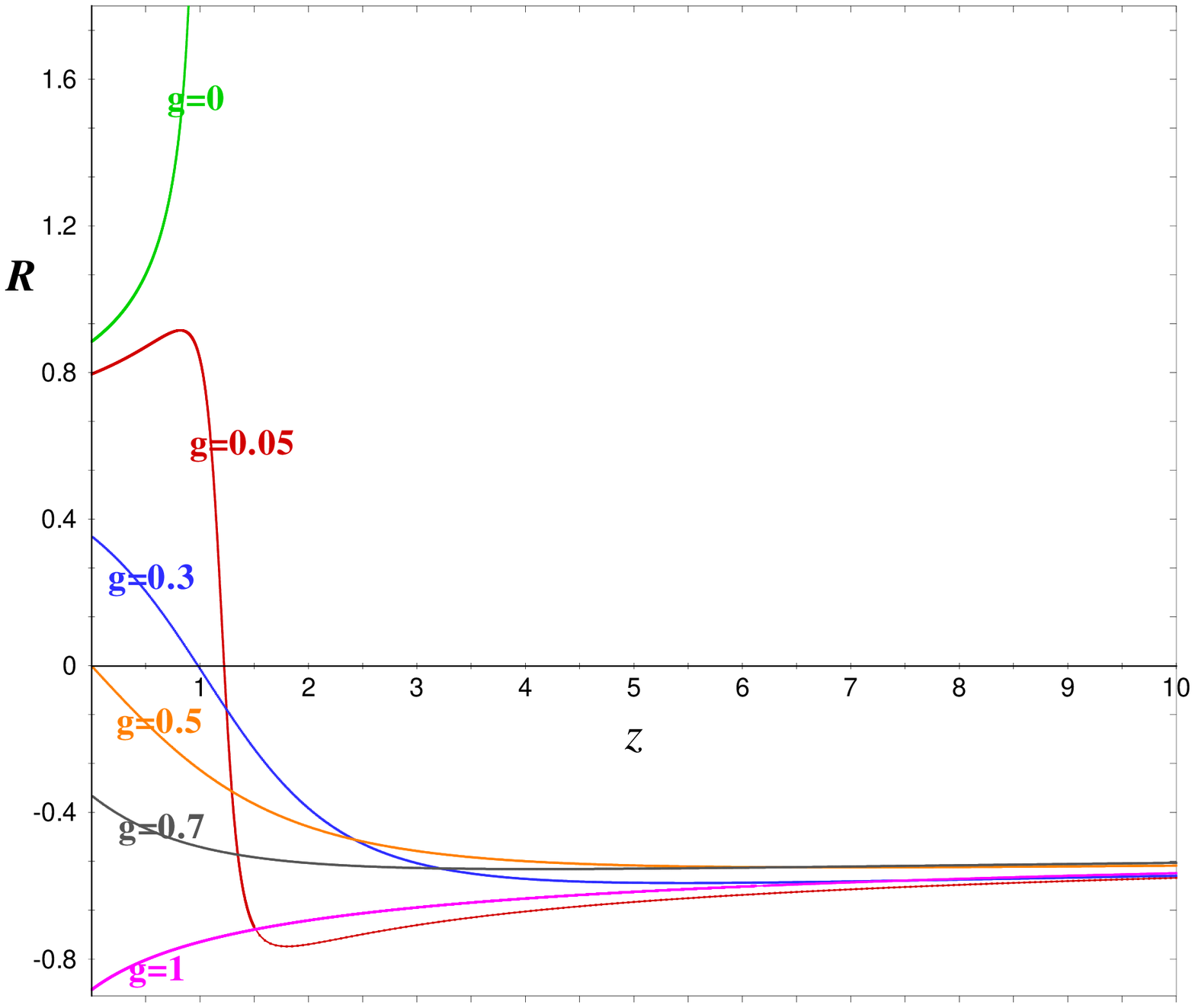}\\
    \caption{(Color online)  The thermodynamic curvature of an ideal 3D  HFE gas
    as a function of fugacity for an isotherm process ($\beta=1$).  The assumed values of the fractional parameter are
     $g=0$ [green (upper) line], 0.05 (red line), 0.3 (blue line), 0.5 (orange line), 0.7 (black line), 1 [purple (lower) line].}\label{figure3}
   \end{figure}

It has been shown that, at T=0, the particles of general exclusion
statistics exhibit a Fermi surface \cite{nayak}. One can observes
that the thermodynamic curvature of the fractional exclusion gas
 tends toward the value
of an ideal fermion gas for any value of fractional parameter
except $g=0$ in the large value of fugacity (low temperature
limit) which is consistent with the result in \cite{nayak}. Also,
there is no singularity for any value of fractional parameter
except for $g=0$ at $z=1$. Hence, there is no phase transition
for $g \neq 0$. It has been shown that there is no phase
transition in the Abelian Haldane fractional exclusion gas
\cite{Geoffrey}.

We restrict ourselves to the region of $z>1$ and investigate the
behavior of  the dual points in different dimensions. Figure
\ref{figure4}. shows the thermodynamic curvature of the HFE gas in
various dimensions and for the same values of the thermodynamic
parameter space.
\begin{figure}[t]
    \center\includegraphics[width=1.1\columnwidth]{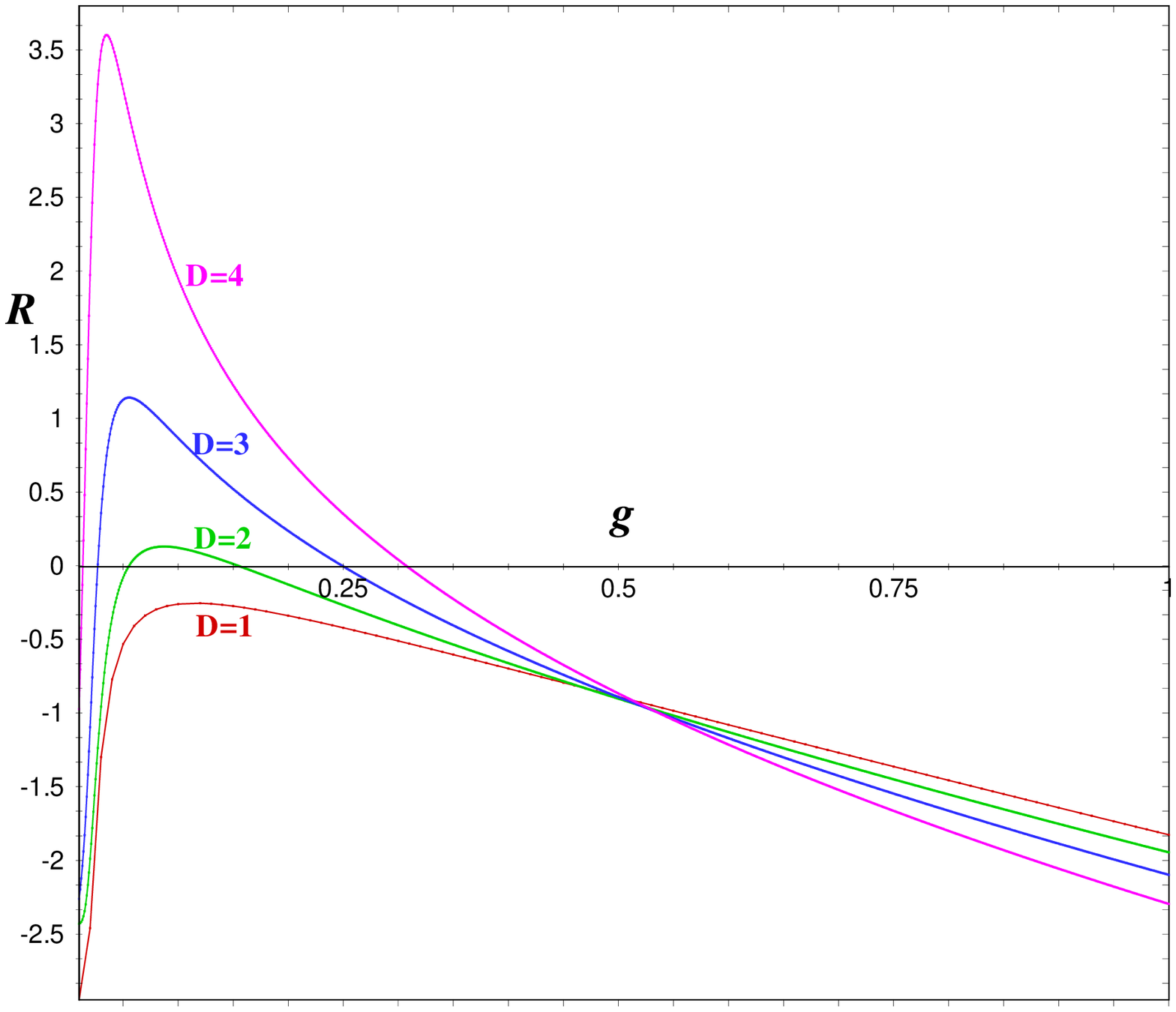}\\
    \caption{(Color online)  The thermodynamic curvature of an ideal HFE gas
    as a function of the fractional parameter  for various dimensions $ D=1, 2, 3, 4$
    at
    a fixed point of the thermodynamic parameters space $\beta=2$ and $z=1.15$.}.\label{figure4}
   \end{figure}

This figure shows that the maximum point goes to the lower value
of fractional parameter in the higher dimension. Also, the dual
points go away from each other  at higher dimensions.  Duality may
also disappear at lower dimensions. The gas is  more stable at
higher dimensions.  Also, we can see from Fig. \ref{figure5} that
for a fixed dimension of space, the dual points are fixed for
different values of temperature. Generally, the sign of the
thermodynamic curvature is independent of $\beta$. In other
words, we can find a threshold value for fugacity ($z_0$) for
which the thermodynamic curvature in $z>z_0$ will be negative for
any value of $\beta$.

\begin{figure}[t]
    \center\includegraphics[width=1.1\columnwidth]{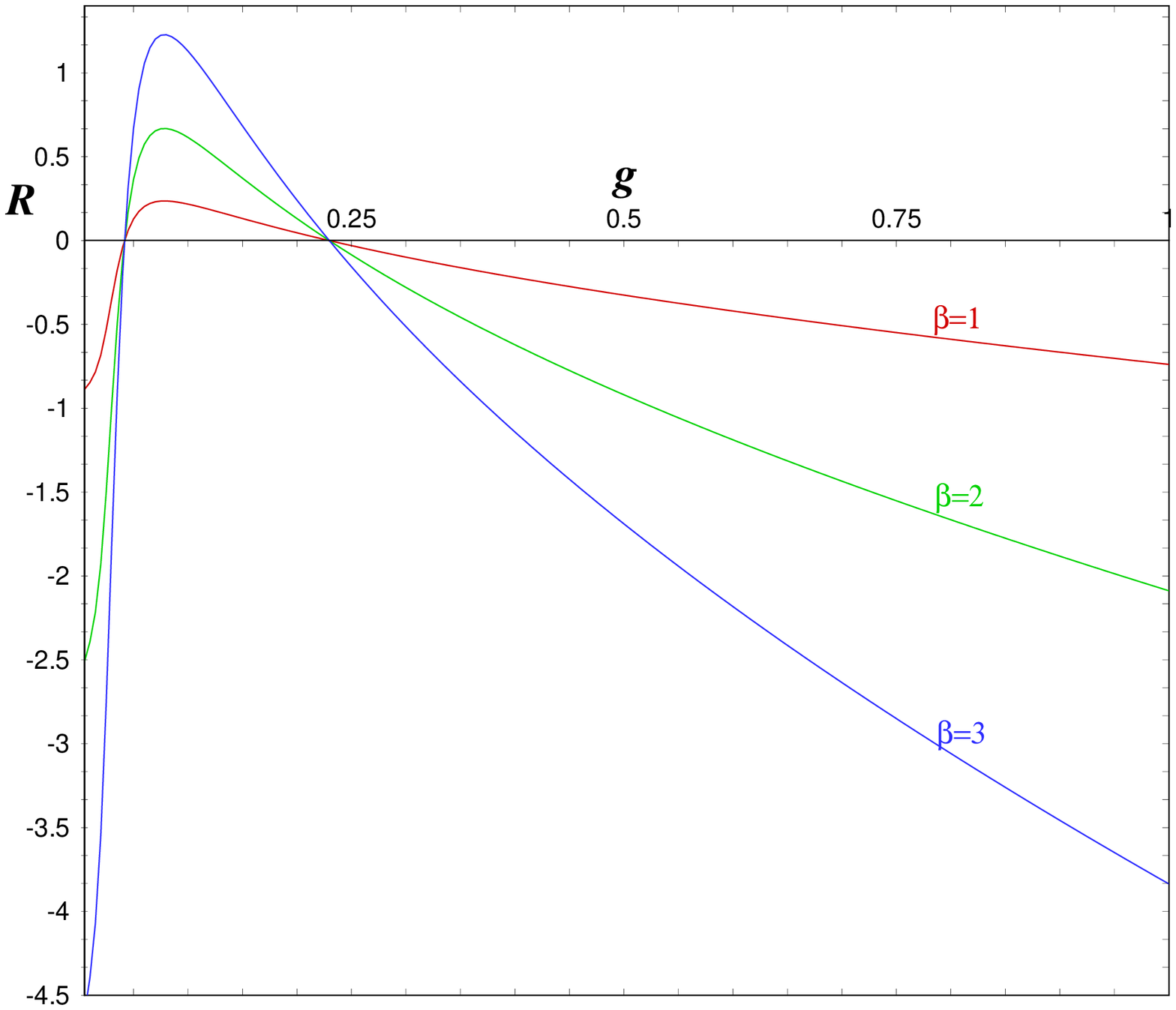}\\
    \caption{(Color online)  The thermodynamic curvature  of an ideal 3D  HFE
    gas (non-relativistic) as a function of the fractional parameter  for a fixed value of fugacity ($z=1.2$)
    and different values of $\beta = 1$ [red (lower max) line], 2 [green (middle max) line], 3 [blue (upper max) line].}\label{figure5}
   \end{figure}

\subsubsection{Ultrarelativistic Limit ($\sigma=1$)}
In this section, we consider the fractional exclusion particle in
the ultra-relativistic limit. First, we represent the result in
two spatial dimensions as in Fig \ref{figure6}.

\begin{figure}[t]
    \center\includegraphics[width=1.2\columnwidth]{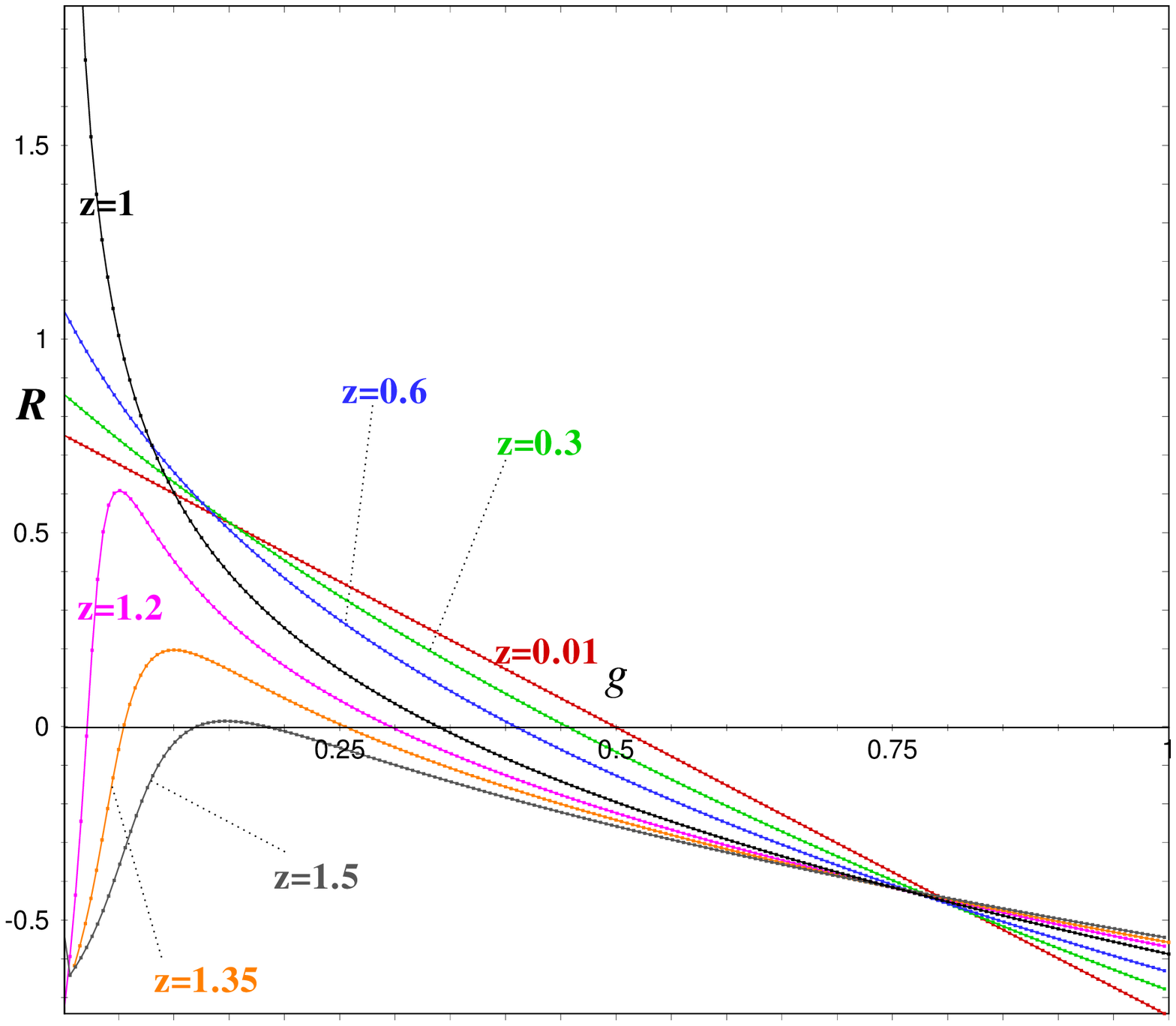}\\
    \caption{(Color online)  The thermodynamic curvature  of a 2D  HFE
    gas (ultra-relativistic limit) as a function of fractional parameter for an isotherm
    ($\beta=1$). The assumed values of anyon fugacity are  $z= 0.01$ (red line) , 0.3 (green line), 0.6 (blue line),
    1 [black (middle) line], 1.2 (purple line), 1.35 (orange line), 1.5 (gray line). }\label{figure6}
   \end{figure}

One can perceive that in the classical limit (small value of
fugacity), the thermodynamic curvature is zero at $g=0.5$, which
is similar to the non-relativistic limit. Also, the general
behavior of the thermodynamic curvature in all its physical range
is similar to its non-relativistic limit.

\begin{figure}[t]
    \center\includegraphics[width=1.2\columnwidth]{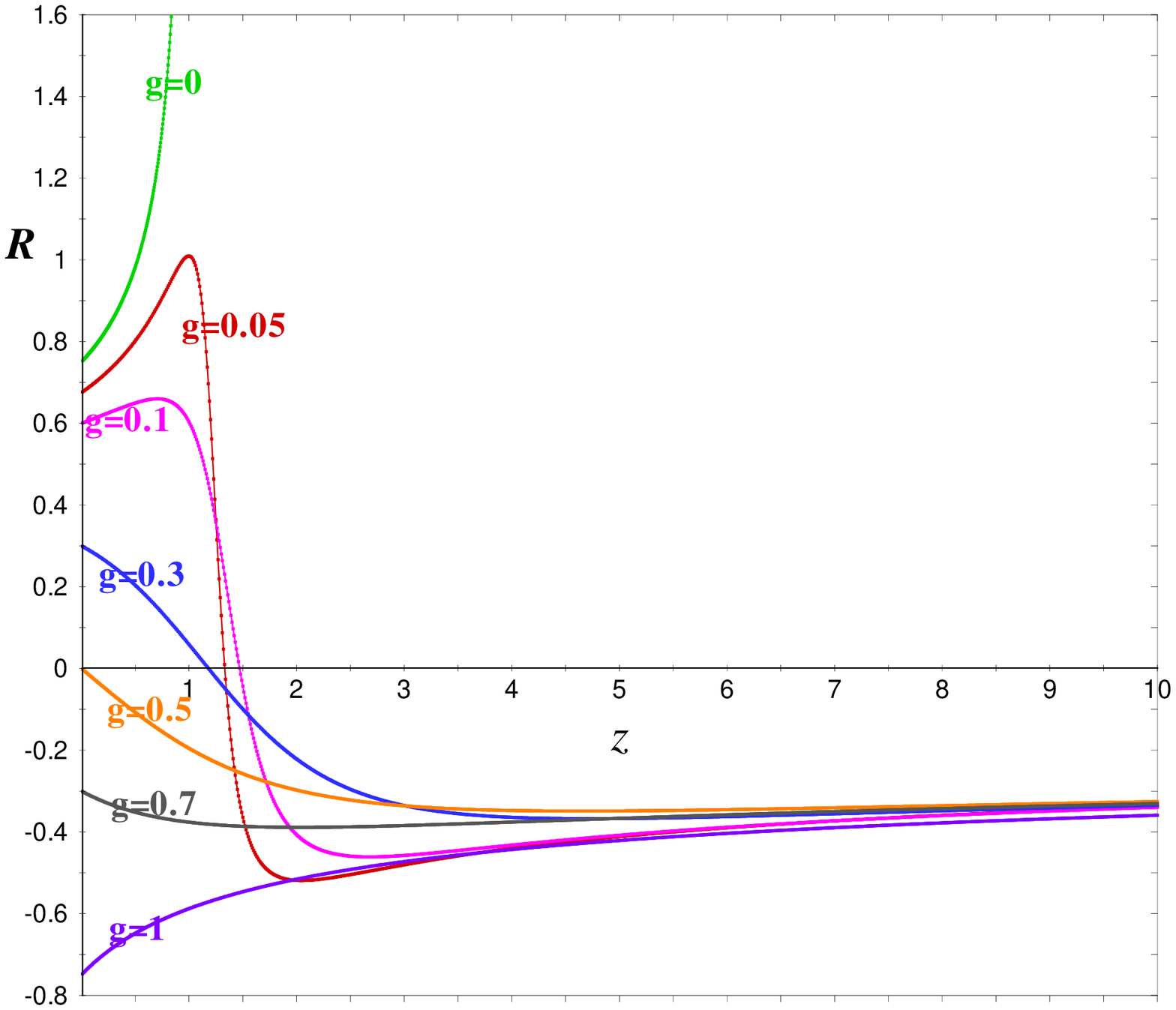}\\
    \caption{(Color online)  The thermodynamic curvature of an
    ideal 2D ((ultra-relativistic) HFE gas  as a function of $z$ for an isotherm
    ($\beta=1$). The assumed values of fractional parameter are $g = 0 $ (green line), 0.05 (red line), 0.1 (purple line), 0.3 (blue line), 0.5 (orange line), 0.7 (black line), 1 (red line).} \label{figure7}
   \end{figure}

The thermodynamic curvature as a function of fugacity and for
various values of fractional parameter is depicted in Fig.
\ref{figure7}. It shows that the thermodynamic curvature of the
boson gas in the ultra-relativistic limit in 2D starts from a
positive value in the classical limit and tends to infinity at
$z=1$. For small values of fractional parameter, the
thermodynamic curvature has a maximum and, finally, goes to the
negative  values  but for large values of the fractional
parameter, it is a monotonically decreasing function with respect
to $z$. A comparison between the non-relativistic and
ultra-relativistic limits shows that the general behavior of the
thermodynamic curvature remains the same. Therefore, it seems that
the dispersion relation does not affect the general behavior of
the thermodynamic curvature and, thereby,the stability of the gas
in different dimensions.

\subsection{Gentile statistics}

The thermodynamic geometry of an ideal gentileonic gas can be
calculated by using the internal energy and the particle number of
the gas. The thermodynamic parameter space is a two dimensional
space with the parameters $(\beta,\gamma)$. Therefore, one can
easily evaluate the metric elements and their derivatives with
respect to the parameters of the thermodynamic space. Thus, we
work out the thermodynamic curvature as a function of
thermodynamic parameters and the fraction parameter $q$. The
thermodynamic curvature of ideal gases obeying Gentile's
statistics has been investigated by Oshima \textit{et al.}
\cite{oshima}. Here, we can consider it in an  arbitrary
dimension and for any value of fractional parameter in the full
physical range. The thermodynamic curvature is depicted in Fig.
\ref{figure8} as a function of fractional parameter, $q$, for
isotherm processes and various values of fugacity, where we
suppose $D=3$ and $\sigma=2$. It is clear that in the classical
limit (small values of fugacity), and in a vast domain of the
fractional parameter, the thermodynamic curvature is mostly
positive and less stable than the HFE gas. Zeroth point of the
ricci scalar goes to the lowest value by increasing the fugacity.
Like the case with the  HFE gas for $z>1$, a maximum point appears
and for some values of the thermodynamic parameter space, we
obtain two values for $q$ with zero curvature, which indicates a
duality relation between these points.

\begin{figure}[t]
    \center\includegraphics[width=0.9\columnwidth]{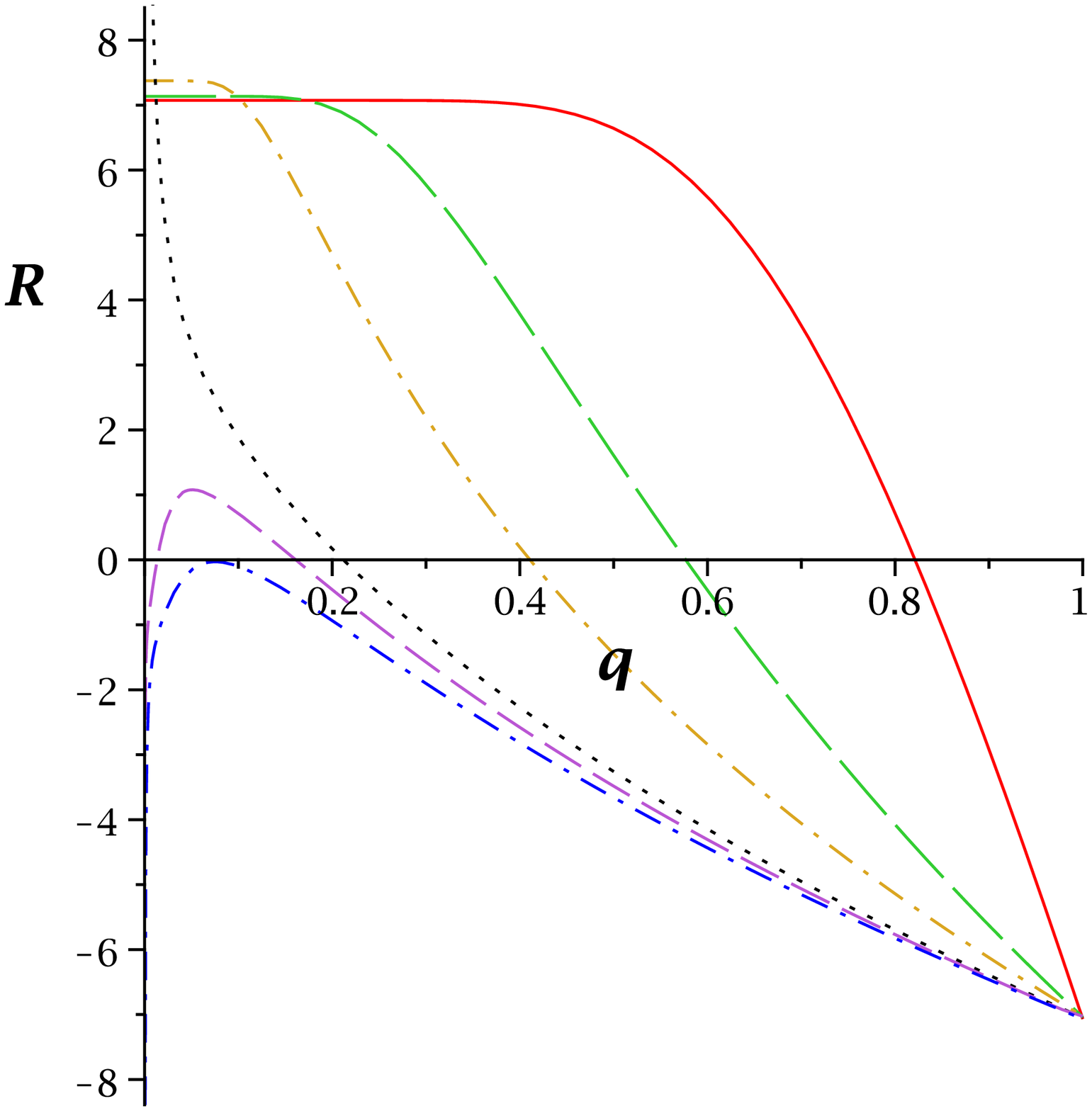}\\
    \caption{(Color online)  The thermodynamic curvature of an ideal
     3D  Gentile statistic gas as a function of $q$ for an isotherm ($\beta=1$). The assumed values of
     fugacity  are  $z=0.01$ [red (solid) line] , 0.2 [green (long dashed) line], 0.6 [brown (dash dotted) line],
      1 [black (dotted) line], 1.2 [violet (dashed) line], 1.3 [blue (lower) line].}\label{figure8}
   \end{figure}

Figure \ref{figure9}. shows that for any value of the fraction
parameter, and for
 large value of fugacity (low temperature limit), the thermodynamic curvature
  tends toward a negative value. Therefore, the particles obeying Gentile statistics, at $T=0$, exhibit a Fermi surfaces.

   \begin{figure}[t]
    \center\includegraphics[width=0.9\columnwidth]{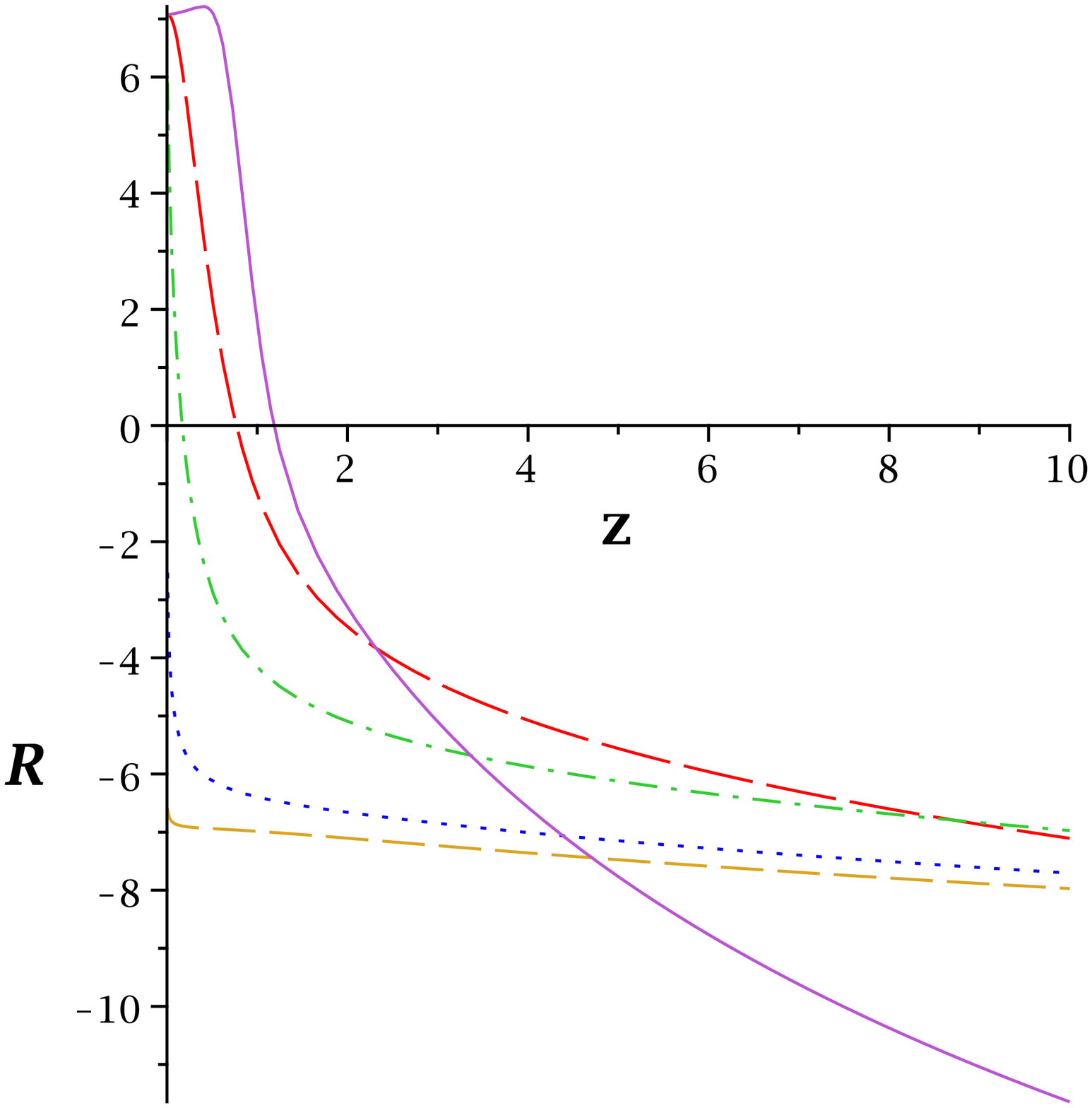}\\
    \caption{(Color online)  The thermodynamic curvature of an ideal  3D
    Gentile statistic gas as a function of fugacity for an isotherm
    ($\beta=1$). The assumed values of the fractional parameter  are $q= 0.1$ [violet (solid) line], 0.3 [red (long dashed) line], 0.6 [green (dash dotted) line], 0.9 [blue (dotted) line],
     1 [brown (dashed) line].}\label{figure9}
   \end{figure}

\subsection{Polychronakos statistics}

Constructing the thermodynamic geometry of an ideal gas with
particles obeying Polychronakos fractional exclusion statistics
is entirely similar to the previous examples. Using Eq.
(\ref{UNP}), we can obtain the  metric elements and the Ricci
scalar curvature in an arbitrary dimension. We will focus on three
spatial dimensions. The thermodynamic curvature is depicted  as a
function of the fractional parameter, $k$, for isotherm processes
in Fig.\ref{figure10} . Various values have been selected in each
curve for fugacity from the classical limit to the quantum limit.
It is obvious that in the high temperature limit, the two kinds of
fractional exclusion statistics, namely Haldane and Polychronakos
fractional exclusion statistics, are the same. In all the physical
range, $k=0.5$ coincides with the zeroth point of the
thermodynamic curvature. It is easy to show that the distribution
function of Polychronakos fractional statistics at $k=0.5$ is the
Boltzmann distribution function. Therefore, an ideal fractional
exclusion gas at $k=0.5$ is an ideal classical gas, which has zero
curvature in the full physical range. Also, for the full physical
range, the thermodynamic curvature is positive for $k<0.5$ but
negative for $k>0.5$.

\begin{figure}[t]
    \center\includegraphics[width=1\columnwidth]{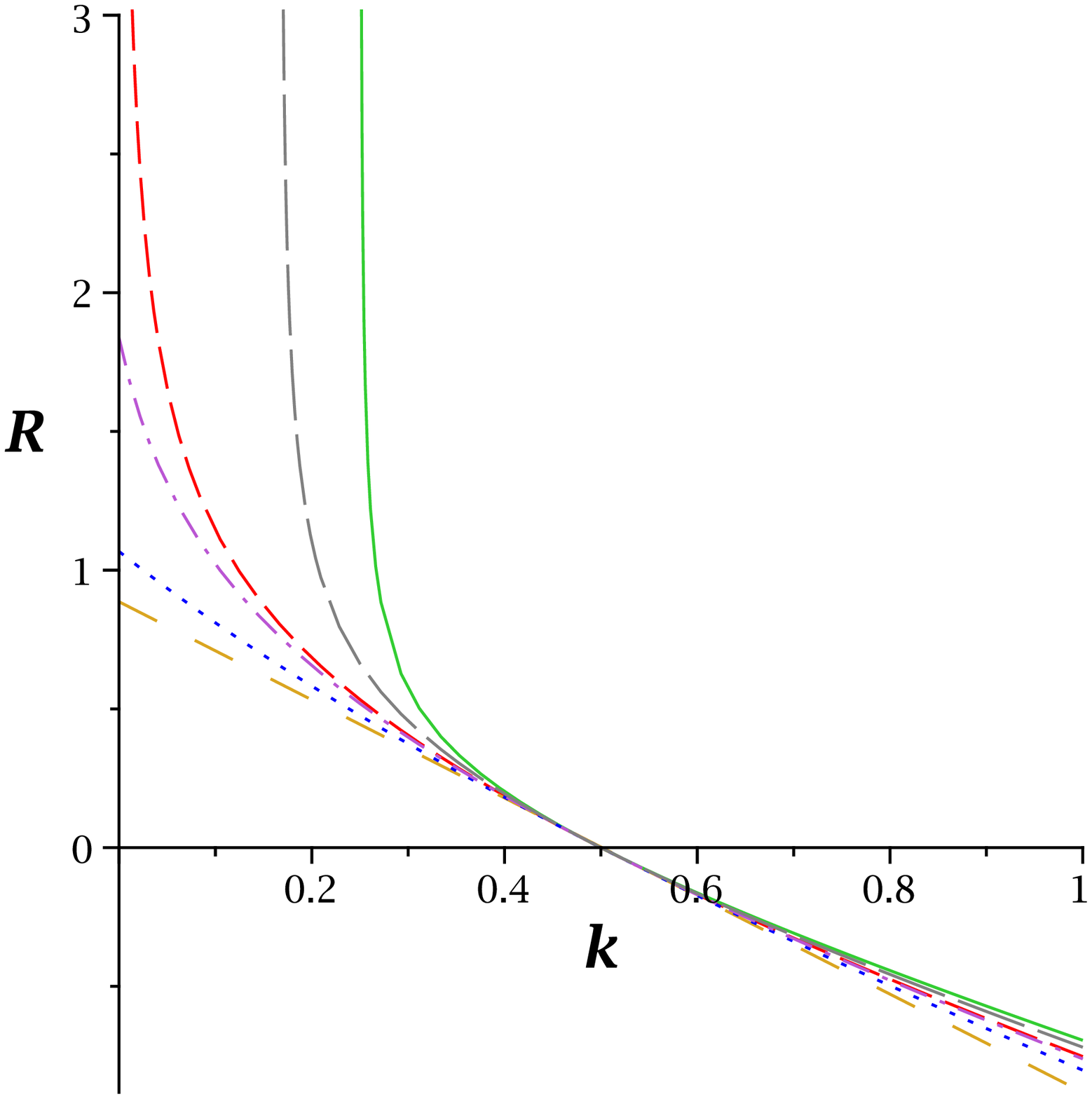}\\
    \caption{(Color online)  The thermodynamic curvature of an
    ideal  3D  PFE gas as a function of $k$ for an isotherm ($\beta=1$). The assumed values
    of fugacity  are $z=2$ [green (solid) line], 1.5 [gray (long dashed) line], 1 [red (dashed) line],
    0.9 [violet (dash dotted) line], 0.5 [blue (dotted) line), 0.01 [brown (lower space dashed) line].}\label{figure10}
   \end{figure}

It has been shown that the thermodynamic curvature of a boson gas
is singular at $z=1$, where
 the Bose-Einstein condensation occurs. This is clear from Figs. \ref{figure10} and \ref{figure11}
 in which
  the thermodynamic curvature goes to infinity for $k=0$ (bosons). However, an interesting new phenomenon is
 observable for $(0<k<0.5)$. For a certain value of the fractional parameter in this interval, there is a specific
 value of fugacity, at which the thermodynamic curvature has a singularity.
 \begin{figure}[t]
    \center\includegraphics[width=1\columnwidth]{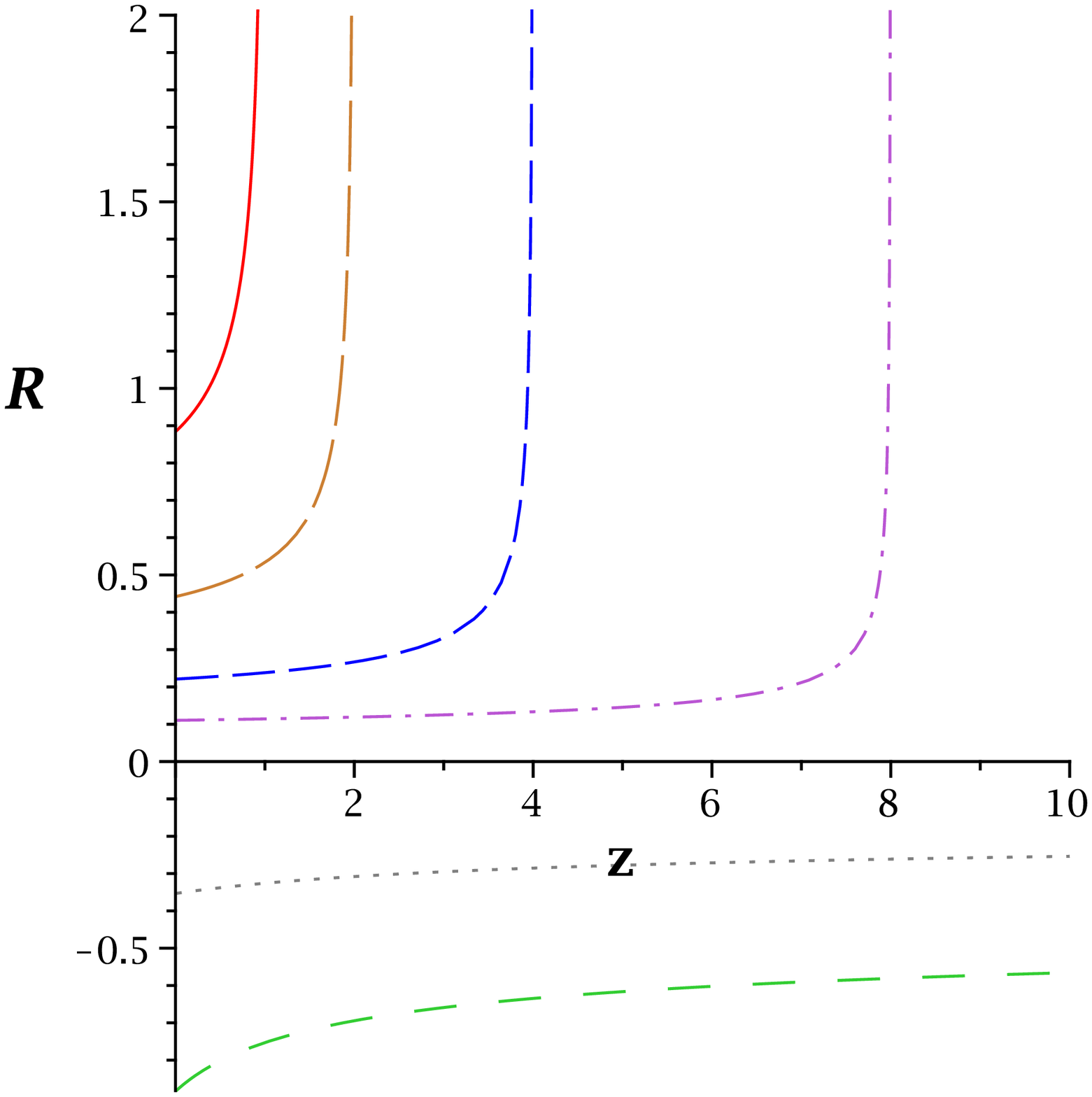}\\
    \caption{(Color online)  The thermodynamic curvature of a  3D  ideal
     PFE gas as a function of fugacity for an isotherm ($\beta=1$). The assumed values
     of the
      fractional parameter are $k = 0$ [red (solid) line],
       $1\over 4$ [brown (long dashed) line], $\frac {3}{8}$ [blue (dashed) line], $\frac {7}{16}$ [violet (dash dotted) line], 0.7 [gray (dotted) line], 1 [green (lower space dashed) line].}\label{figure11}
   \end{figure}

\section{Condensation for Particles with Fractional Statistics}

Condensation of particles obeying Polychronakos fractional
statistics for different values of the fractional parameter is an
interesting subject. This kind of condensation $(R=\infty)$
appears for non-pure bosonic particles. In fact, the singularity
of the thermodynamic curvature arises from the definition of the
polylogarithm function ($Li_{n}(x)$), where $x$ has to belong to
the interval $]-\infty,1]$. Thereupon, one can shows from the
thermodynamic quantities such as internal energy and particle
number that where $z-2kz=1$, the thermodynamic curvature face
with a singularity. Therefore, for a given value of the
fractional parameter, $k$, the singularity occurs at

 \bea
 z_{c}=\frac{1}{1-2k}.\label{wq}
 \eea
In fact, it indicates a condensation similar to the Bose-Einstein
but for a non-zero fractional parameter or a non-pure bosonic
system of the fractional gas. Thermodynamic curvature for boson gas and fractional exclusion statistic gas is a function with respect to $\beta$ and $z$ in the following general form
 \bea
 R=\beta^{\frac{D}{\sigma}}f(z,k,D,\sigma).
 \eea
If we set $z=1$ (boson case) and $z=1/(1-2k)$ (Polychronakos statistics), the thermodynamic curvature goes to infinity independent of $\beta$. Therefore, the thermodynamic curvature diverges at critical fugacity for all value of $\beta$. For obtaining the critical temperature we need an additional relation such as the particle number.
By substituting $a= {1\over 2m}$ and
$\sigma = 2$,  Eq. (\ref{UNP}) yields
 \bea
 n\lambda^{3}(1-2k)={Li_{\frac{3}{2}}(z-2kz)},\label{TC}
 \eea

\noindent where, $n=\frac{N}{L^3}$ and
$\lambda=\sqrt{\frac{h^2}{2m\pi k_{_B} T}}$ is the
 thermal wavelength. For a given value of $k$, there is a suitable value
  for fugacity that satisfies  Eq. (\ref{wq}) and $Li_{\frac{3}{2}}(1)=\zeta(\frac{3}{2})$ has
  a finite value. However, it is obvious from Fig. \ref{figure12} that for
  the  values $k=\frac{1}{4},\frac{1}{3},\frac{3}{8},\cdots$, the function $n\lambda^{3}(1-2k)$
  with
  an infinite slope at $z=2,3,4,\ldots$ which is similar to  an ideal boson gas that  has infinite slope at $z=1$
  where the Bose-Einstein condensation occurs \cite{Huang}.

  \begin{figure}[t]
    \center\includegraphics[width=1\columnwidth]{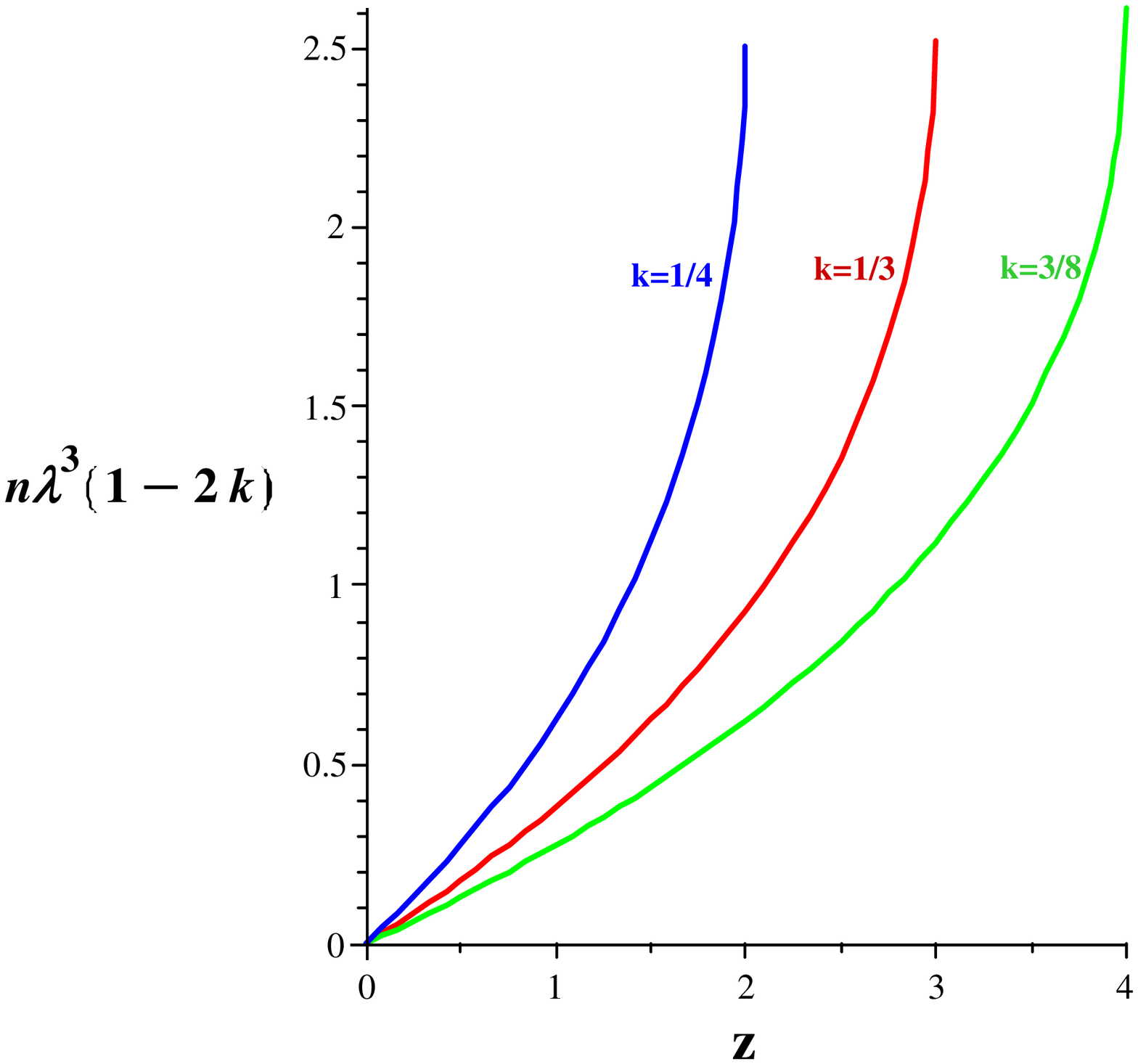}\\
    \caption{(Color online)\  $n\lambda^{3}(1-2k)$ as a function of
     fugacity for some values of fractional parameter, $k=\frac {1}{4}$ (blue line), $\frac {1}{3}$ (red line), $\frac {3}{8}$  (green line). }\label{figure12}
   \end{figure}

\noindent In constant $n$, we can introduce a critical temperature
from Eq. (\ref{TC}),
 \bea
 k_{_B}{T}_{c}=\frac{h^2}{2m\pi}[\frac{n(1-2k)}{\zeta(3/2)}]^{2/3},
 \eea
where, $\zeta(x)$ is the Riemann zeta function. Also, one can
find a restricting condition on the chemical potential of
particles with the
 fraction parameter $k$. Whereas $z$ has to satisfy the inequality $z\leq\frac{1}{1-2k}$,  we obtain
 \bea
 \mu_{k}\leq-k_{_B}T\ln(1-2k).
 \eea

\noindent We can find the transition temperature in an arbitrary
dimension and  with a general dispersion relation,

 \bea
 n\Lambda^{\frac{D}{\sigma}}(1-2k)=\frac{\Gamma(\frac{D}{\sigma})}{\Gamma(\frac{D}{2})}Li_{\frac{D}{\sigma}}(z-2kz),
 \eea
where, $n=\frac{N}{L^{D}}$ is the density of the particles in an
arbitrary dimension and
$\Lambda=\frac{ah^{\sigma}}{\pi^{\frac{\sigma}{2}}k_{B}T}$. Now,
 similar to the 3D case,  condensation occurs for $z=\frac{1}{1-2k}.$ Thus, the transition temperature in constant $n$ is given by
 \bea
 k_{B}T_{c}=\frac{a h^{\sigma}}{\pi^{\sigma/2}}\left(\frac{n(1-2k)\Gamma(\frac{D}{2})}{\Gamma(\frac{D}{\sigma})\zeta(\frac{D}{\sigma})}\right)^{\sigma/D}.
 \eea
According to the zeta function properties, it is obvious that
there is a finite transition temperature for
$\frac{D}{\sigma}>1$. Furthermore, we notice that in all
thermodynamic quantities and all metric elements of the
thermodynamic geometry, the dependency on $D$ and $\sigma$ appears
as $\frac{D}{\sigma}$. Therefore, the
 thermodynamic properties such as critical properties for $D=2n$ dimensional fractional statistic gas in the non-relativistic
 limit are similar to the $D=n$ dimensional one in the ultra-relativistic limit \cite{May}. In other words,
  condensation occurs  in
$D\geq 3$ in the non-relativistic limit  but in $D\geq 2$ in the
ultra-relativistic limit.

\section{Conclusion}

The internal energy and particle number of three kinds of
fractional statistics were represented in an arbitrary  dimension.
The thermodynamic geometry of an ideal Haldane, Gentile and
Polychronakos fractional exclusion statistic gas was also explored
in arbitrary dimensions. Finally, the thermodynamic curvature of
Haldane statistics was worked out in the non-relativistic and
ultra-relativistic limits in 2D and 3D. It was found that in the
non-relativistic limit, the results for 3D are  similar to those
in the 2D. However, moving to higher dimensions, the
thermodynamic geometry suggested that less stable gases would be
obtained. It was also found that  no singularity exists for any
value of the  fractional parameter, except for $g=0$ at $z=1$,
which corresponds to the Bose-Einstein condensation.

\noindent The thermodynamic geometry of an ideal gas obeying
 Gentile statistic was considered. The thermodynamic curvature
 in the classical limit was found to be  mostly  positive, which may be related to the
 attractive statistical interaction of Gentileons. Moving from the classical limit to the larger values of fugacity at $z>1$, one
 can observe a similarity between the  HFE gas and  the Gentileonic gas. For some values
  of thermodynamic parameters, there is a duality between two different values of
  the fractional parameter in both cases. Also, it seems that at $T=0$, particles
  obeying the
  HFE statistics and Gentile statistics exhibit a Fermi surfaces and the repulsive
  statistical interaction is dominant in the low temperature limit. Again,
  only the singular point in a  gentileonic gas is related to  $q=0$ (boson) at
  $z=1$. In general, the HFE gas is more stable than the Gentileonic gas.

\noindent Some special phenomena were found to occur for an ideal
Polychronakos fractional exclusion gas. One can find that in the
classical limit, two different definitions of the fractional
exclusion statistics, namely Haldane and Polychronakos, coincide. It is mentionable that these statistics
are different completely in full physical range.  Specially, in low temperature limit the exclusion property
of particles obeying Haldane statistics is dominant. Therefore,  BEC does not occur for particles obeying HFE statistics.
In other words, Polychronakos statistics does not drive from Haldane statistics, but these statistics coincide
with together in the classical limit. Going far away from the classical limit, their differences begin
to appear. The thermodynamic geometry of a PFE gas can be divided
into two different regions. For $k<0.5$, the thermodynamic
curvature is positive (Bose-like), while for $k>0.5$, it is
negative (Fermi-like). Also, there are  some singular points of
thermodynamic curvature for $k<0.5$ (Bose-like) which correspond
to the condensation of the fractional statistics particles. Thus,
we find a condensation for a non-pure bosonic system.  In all
kinds of fractional statistics, a non-relativistic
$2d$-dimensional fractional statistics gas  has the same
thermodynamic geometry that an ultra-relativistic gas in
$d$-dimensions.


\end{document}